\documentclass[reprint,twocolumn,aps,prc,a4paper, superscriptaddress,showpacs,preprintnumbers]{revtex4}
\usepackage{amssymb}
\usepackage{amsmath}
\usepackage{multirow}
\usepackage{braket}
\usepackage{graphicx}
\usepackage{subfigure}
\usepackage{lscape}
\usepackage{booktabs,longtable}
\usepackage{mathptmx, courier, pifont}
\usepackage[scaled=0.92]{helvet}
\usepackage[T1]{fontenc}
\usepackage{textcomp}
\usepackage[dvipsnames]{xcolor}
\usepackage{verbatim}
\usepackage{amsmath}

\newcommand{\iso}[2]{$^{#1}$#2}
\newcommand{\orb}[3]{#1#2$_{#3/2 }$}
\newcommand{\ch}[2]{(\iso{#1}{F},\iso{#2}{O})}

\definecolor{deepblue}{RGB}{0,32,96}
\definecolor{green}{RGB}{0,176,80}
\definecolor{orange}{RGB}{248,165,96}
\definecolor{lightblue}{RGB}{0,112,192}
\definecolor{lightgreen}{RGB}{155,187,89}
\definecolor{deepred}{RGB}{119,44,42}
\definecolor{purple}{RGB}{112,48,160}
\definecolor{olivegreen}{RGB}{0,50,0}
\definecolor{lightblue}{RGB}{28,94,255}
\definecolor{deepviolet}{RGB}{58,00,111}

\begin{document}

\title{How different is the core of \iso{25}{F} from \iso{24}{O}$_{\textrm{g.s.}}$?}

%\begin{comment}
\author{T. L. Tang}
\thanks{ttang@anl.gov; Present address: Physics, Division, Bldg 203, Argonne National Laboratory, 9700 South Cass Avenue, Lemont, IL 60439, USA}
\affiliation{Center for Nuclear Study, University of Tokyo, 7-3-1 Hongo, Bunkyo, Tokyo 113-0033, Japan}
\affiliation{RIKEN Nishina Center, 2-1 Hirosawa, Wako, Saitama 351-0198, Japan}
\author{T. Uesaka}\affiliation{RIKEN Nishina Center, 2-1 Hirosawa, Wako, Saitama 351-0198, Japan}
\author{S. Kawase}
\thanks{Present address: Nuclear Science and Engineering Center, Japan Atomic Energy Agency,
Tokai, Ibaraki 319-1195, Japan}\affiliation{Center for Nuclear Study, University of Tokyo, 7-3-1 Hongo, Bunkyo, Tokyo 113-0033, Japan}
\author{D. Beaumel}\affiliation{Institut de physique nucl\'eaire d$'$Orsay, 91406 Orsay, France}
\author{M. Dozono}\affiliation{RIKEN Nishina Center, 2-1 Hirosawa, Wako, Saitama 351-0198, Japan}
\author{T. Fujii}\affiliation{Center for Nuclear Study, University of Tokyo, 7-3-1 Hongo, Bunkyo, Tokyo 113-0033, Japan}
\author{N. Fukuda}\affiliation{RIKEN Nishina Center, 2-1 Hirosawa, Wako, Saitama 351-0198, Japan}
\author{T. Fukunaga}\affiliation{Kyushu University, 6-10-1 Hakozaki, Higashi, Fukuoka 812-8581, Japan}
\author{A. Galindo-Uribarri}\affiliation{Oak Ridge National Laboratory, 1 Bethel Valley Rd, Oak Ridge, TN 37831, USA}
\author{S. H. Hwang}\thanks{Present  addrress: Korea Research Institute of Standards and Science, Daejeon 34113, Republic of Korea}\affiliation{Kyungpook National University, 80 Daehakro, Bukgu, Daegu, 41566, Republic of Korea}%shhwang@kriss.re.kr; 
\author{N. Inabe}\affiliation{RIKEN Nishina Center, 2-1 Hirosawa, Wako, Saitama 351-0198, Japan}
\author{D.~Kameda}\affiliation{RIKEN Nishina Center, 2-1 Hirosawa, Wako, Saitama 351-0198, Japan}
\author{T.~Kawahara}\affiliation{Toho University, 2-2-1 Miyama, Funabashi-shi, Chiba 274-8510, Japan} 
\author{W.~Kim}\affiliation{Kyungpook National University, 80 Daehakro, Bukgu, Daegu, 41566, Republic of Korea}
\author{K.~Kisamori}\affiliation{Center for Nuclear Study, University of Tokyo, 7-3-1 Hongo, Bunkyo, Tokyo 113-0033, Japan}
\author{M.~Kobayashi}\affiliation{Center for Nuclear Study, University of Tokyo, 7-3-1 Hongo, Bunkyo, Tokyo 113-0033, Japan} 
\author{T.~Kubo}\affiliation{RIKEN Nishina Center, 2-1 Hirosawa, Wako, Saitama 351-0198, Japan}
\author{Y.~Kubota}\thanks{Present address: Institut f\"ur Kernphysik, Technische Universit\"at Darmstadt, 64289 Darmstadt, Germany}\affiliation{Center for Nuclear Study, University of Tokyo, 7-3-1 Hongo, Bunkyo, Tokyo 113-0033, Japan}
\author{K.~Kusaka}\affiliation{RIKEN Nishina Center, 2-1 Hirosawa, Wako, Saitama 351-0198, Japan}
\author{C.~S.~Lee}\affiliation{Center for Nuclear Study, University of Tokyo, 7-3-1 Hongo, Bunkyo, Tokyo 113-0033, Japan}
\author{Y.~Maeda}\affiliation{University of Miyazaki, 1-1 Gakuen Kibanadai-nishi, Miyazaki 889-2192, Japan}
\author{H.~Matsubara}\thanks{Present address: Graduate School of Medical Science, Tokyo Women\textquotesingle s Medical University,  Tokyo, 162-8666, Japan}\affiliation{RIKEN Nishina Center, 2-1 Hirosawa, Wako, Saitama 351-0198, Japan}
\author{S.~Michimasa}\affiliation{Center for Nuclear Study, University of Tokyo, 7-3-1 Hongo, Bunkyo, Tokyo 113-0033, Japan}
\author{H.~Miya}\affiliation{Center for Nuclear Study, University of Tokyo, 7-3-1 Hongo, Bunkyo, Tokyo 113-0033, Japan}
\author{T.~Noro}\affiliation{Kyushu University, 6-10-1 Hakozaki, Higashi, Fukuoka 812-8581, Japan} 
\author{A.~Obertelli}\thanks{Present address: Institut f\"ur Kernphysik, Technische Universit\"at Darmstadt, 64289 Darmstadt, Germany}\affiliation{RIKEN Nishina Center, 2-1 Hirosawa, Wako, Saitama 351-0198, Japan}\affiliation{IRFU, CEA, Universit\'e Paris-Saclay, 91191 Gif-sur-Yvette, France}
\author{K.~Ogata}\affiliation{RCNP, Osaka University, 10-1 Mihogaoka, Ibaraki, Osaka, 567-0047, Japan}
\affiliation{Department of Physics, Osaka City University, Osaka 558-8585, Japan}
\author{S.~Ota}\affiliation{Center for Nuclear Study, University of Tokyo, 7-3-1 Hongo, Bunkyo, Tokyo 113-0033, Japan}
\author{E.~Padilla-Rodal}\affiliation{Universidad Nacional Aut\'{o}noma de M\'{e}xico, Instituto de Ciencias Nucleares, AP 70-543, M\'{e}xico City 04510, DF, M\'{e}xico}
\author{S.~Sakaguchi}\affiliation{Kyushu University, 6-10-1 Hakozaki, Higashi, Fukuoka 812-8581, Japan} 
\author{H.~Sakai}\affiliation{RIKEN Nishina Center, 2-1 Hirosawa, Wako, Saitama 351-0198, Japan}
\author{M.~Sasano}\affiliation{RIKEN Nishina Center, 2-1 Hirosawa, Wako, Saitama 351-0198, Japan}
\author{S.~Shimoura}\affiliation{Center for Nuclear Study, University of Tokyo, 7-3-1 Hongo, Bunkyo, Tokyo 113-0033, Japan}
\author{S.~S.~Stepanyan}\affiliation{Kyungpook National University, 80 Daehakro, Bukgu, Daegu, 41566, Republic of Korea} 
\author{H.~Suzuki}\affiliation{RIKEN Nishina Center, 2-1 Hirosawa, Wako, Saitama 351-0198, Japan}
\author{M.~Takaki}\affiliation{Center for Nuclear Study, University of Tokyo, 7-3-1 Hongo, Bunkyo, Tokyo 113-0033, Japan}
\author{H.~Takeda}\affiliation{RIKEN Nishina Center, 2-1 Hirosawa, Wako, Saitama 351-0198, Japan}
\author{H.~Tokieda}\affiliation{Center for Nuclear Study, University of Tokyo, 7-3-1 Hongo, Bunkyo, Tokyo 113-0033, Japan}
\author{T.~Wakasa}\affiliation{Kyushu University, 6-10-1 Hakozaki, Higashi, Fukuoka 812-8581, Japan}
\author{T.~Wakui}\thanks{Present address: National Institute of Radiological Sciences, National Institutes for Quantum and Radiological Science and Technology, Chiba 263-8555, Japan}\affiliation{CYRIC, Tohoku University, 6-3 Aoba, Aramaki, Aoba-ku, Sendai, Miyagi, 980-8578, Japan}
\author{K.~Yako}\affiliation{Center for Nuclear Study, University of Tokyo, 7-3-1 Hongo, Bunkyo, Tokyo 113-0033, Japan}
\author{Y.~Yanagisawa}\affiliation{RIKEN Nishina Center, 2-1 Hirosawa, Wako, Saitama 351-0198, Japan}
\author{J.~Yasuda}\affiliation{Kyushu University, 6-10-1 Hakozaki, Higashi, Fukuoka 812-8581, Japan}
\author{R.~Yokoyama}\affiliation{Center for Nuclear Study, University of Tokyo, 7-3-1 Hongo, Bunkyo, Tokyo 113-0033, Japan} 
\author{K.~Yoshida}\affiliation{RIKEN Nishina Center, 2-1 Hirosawa, Wako, Saitama 351-0198, Japan}
\author{K.~Yoshida}\thanks{Present address: Advanced Science Research Center, Japan Atomic Energy Agency, Tokai, Ibaraki 319-1195, Japan}\affiliation{RCNP, Osaka University, 10-1 Mihogaoka, Ibaraki, Osaka, 567-0047, Japan}
\author{J.~Zenihiro}\affiliation{RIKEN Nishina Center, 2-1 Hirosawa, Wako, Saitama 351-0198, Japan}  

%\end{comment}

\begin{abstract}
The neutron-shell structure of \iso{25}{F} was studied using quasi-free (\textit{p},2\textit{p}) knockout reaction at \mbox{270\textit{A} MeV} in inverse kinematics. The sum of spectroscopic factors of $\pi$\orb{0}{d}{5} orbital is found to be $1.0\pm0.3$. However, the spectroscopic factor for the ground-state to ground-state transition (\iso{25}{F}, \iso{24}{O}$_\textrm{g.s.}$) is only $0.36\pm0.13$, and \iso{24}{O} excited states are produced from the \orb{0}{d}{5} proton knockout. The result shows that the \iso{24}{O} core of \iso{25}{F} nucleus significantly differs from a free \iso{24}{O} nucleus, and the core consists of $\sim$35\% \iso{24}{O}$_\textrm{g.s.}$ and $\sim$65\% excited \iso{24}{O}. 

\end{abstract}

%\pacs{25.60.Gc, 25.70.Hi, 25.70.Mn, 25.70.Pq}
\maketitle

In the independent particle picture \cite{Vijay97}, a nucleon is moving without any correlation with other nucleons. Building upon this, shell models successfully describe many nuclear properties by introducing residual interactions~\cite{Caurier05}. The residual interactions perturb the single particle wavefunction and cause nucleons to become correlated. Theoretically, the degree of independence of a nucleon of a specific single particle orbital (SPO) can be characterized using the spectroscopic factor (SF)~\cite{Glendenning04}, which is related to the occupancy of the SPO. Using electron or proton-induced knockout or transfer reactions on most stable nuclei, the integrated SFs for a given orbital are limited to $\sim0.7$~\cite{Kramer01, Wakasa17}. The reduction from unity is attributed to the short- and long-range correlations among the nucleons~\cite{Lapikas93}.

In the particular case of nuclei with a single valence nucleon and a doubly magic core, the residual interaction between the valence nucleon and the core is weak due to the closed-shell nature of the core. For example, the pairing correlation is weak~\cite{Frauendorf14}.~The SFs of ground-state to ground-state transitions are experimentally in the range of \mbox{$0.8-1.1$} for \mbox{$^{17}\textrm{F}=\textrm{p}+^{16}$O$_\textrm{g.s}$}~\mbox{\cite{Oliver69,Fortune75,Yasue92}}, \mbox{$^{41}\textrm{Sc}=$ p$+^{40}$Ca$_\textrm{g.s}$}~\cite{Leighton68,Youngblood68}, \mbox{$^{49}\textrm{Sc}=$ p$+^{48}$Ca$_\textrm{g.s}$}~\cite{Britton76}, and \mbox{$^{209}\textrm{Bi}=$ p$+^{208}$Pb$_\textrm{g.s}$}~\cite{Berthier86,Gales84,Branford00}. 

A (\textit{p},2\textit{p}) quasi-free knockout reaction can be used to extract the SF of a proton SPO \mbox{($\pi$-SPO)}~\cite{Glendenning04,Jacob66} using experimental cross sections and theoretical single-particle cross sections. With the approximations in Ref.~\cite{Glendenning04,Jacob66}, the cross section ($\sigma$) is proportional to the square of the overlap between a target (\iso{A}{Z}) and the knocked-out proton and a residual nucleus [\iso{A-1}{(Z-1)}] wave functions~\cite{Glendenning04} as
\begin{equation}\label{eq:0}
\sigma\propto\left|\left\lbrace\langle\pi_j|\otimes\langle^{A-1}{(Z-1)^i}|\right\rbrace | ^{A}Z \rangle \right|^2 = \alpha^i(\nu)\beta^i(\pi_j),
\end{equation}
where $|\pi_j\rangle$ is the $j$-th $\pi$-SPO, $\otimes$ is the spin--isospin coupling and antisymmetrization operator. $i$ indicates the state of the residual nucleus. $\alpha^i(\nu)$ is the overlap of the neutron orbitals and $\beta^i(\pi_j)$ is the overlap of the proton orbitals. Usually, in proton-removal reactions such as \iso{16}{O}(\textit{p},2\textit{p})~\cite{McDonald86} and \iso{40}{Ca}(\textit{p},2\textit{p})~\cite{Yasuda10}, the neutron part is assumed to be inert and $\alpha^i(\nu)=1$. Thus the SF is directly related to the overlap of the proton orbitals. 

In the special case where the knocked-out proton is in a SPO to a good approximation, i.e., $\beta^j(\pi_j)=1$, the reaction can be used to investigate the neutron-shell structure via the overlap $\alpha^i(\nu)$. In this Letter we apply this method to neutron-rich fluorine isotopes of \iso{23,25}{F}. The cores of \iso{23}{F} and \iso{25}{F} are \iso{22}{O} and \iso{24}{O} with a proton magic number of 8 and neutron semi-magic number of 14 and 16~\cite{Ozawa00,Dominguez11}, respectively. Particularly, \iso{24}{O} is known to have established doubly magic nature~\cite{Nociforo09, Janssens09}. The $pn$ correlation energies of \iso{23}{F} and \iso{25}{F} are small and equal \mbox{0.7 MeV} and \mbox{0.1 MeV}, respectively~\cite{Kaneko04}. Therefore, the $pn$ pairing correlation is considered to be weak. In addition, the oxygen dripline anomaly, i.e., the drastic change in the neutron dripline from $N=16$ for oxygen to $N=22$ for fluorine, indicates that the structure of the neutron shell may drastically change due to the addition of one \orb{0}{d}{5} proton. A previous work reports that the SF extracted from the $^{12}$C\ch{25}{24} reaction at 50.4\textit{A}~MeV is as small as 0.4~\cite{Thoennessen03}. No detailed discussion is made in Ref.~\cite{Thoennessen03} and it is concluded that experiments with higher precision are necessary in order to infer a change of the structure. On the other hand, the (\textit{p},2\textit{p}) reaction has demonstrated to be a good spectroscopic tool to investigate single particle properties of stable~\cite{Wakasa17} and unstable nuclei~\cite{Kawase18,Atar18}. 

In this letter, the experimental results for a (\textit{p},2\textit{p}) proton knockout reaction of \iso{23,25}{F} are presented. The \iso{25}{F} data is compared to the  distorted-wave impulse approximation (DWIA) calculations. The overlap between the oxygen core of \iso{25}{F} and the \iso{24}{O}$_\textrm{g.s.}$ is as small as $\sim$35\%. The data for \iso{23}{F} shows similar feature.

%===============  fig. 1  ==============
\begin{figure}[!htb]
\centering
\includegraphics[trim={0.5cm 3cm 1cm 5cm}, clip=true, width=8.5cm]{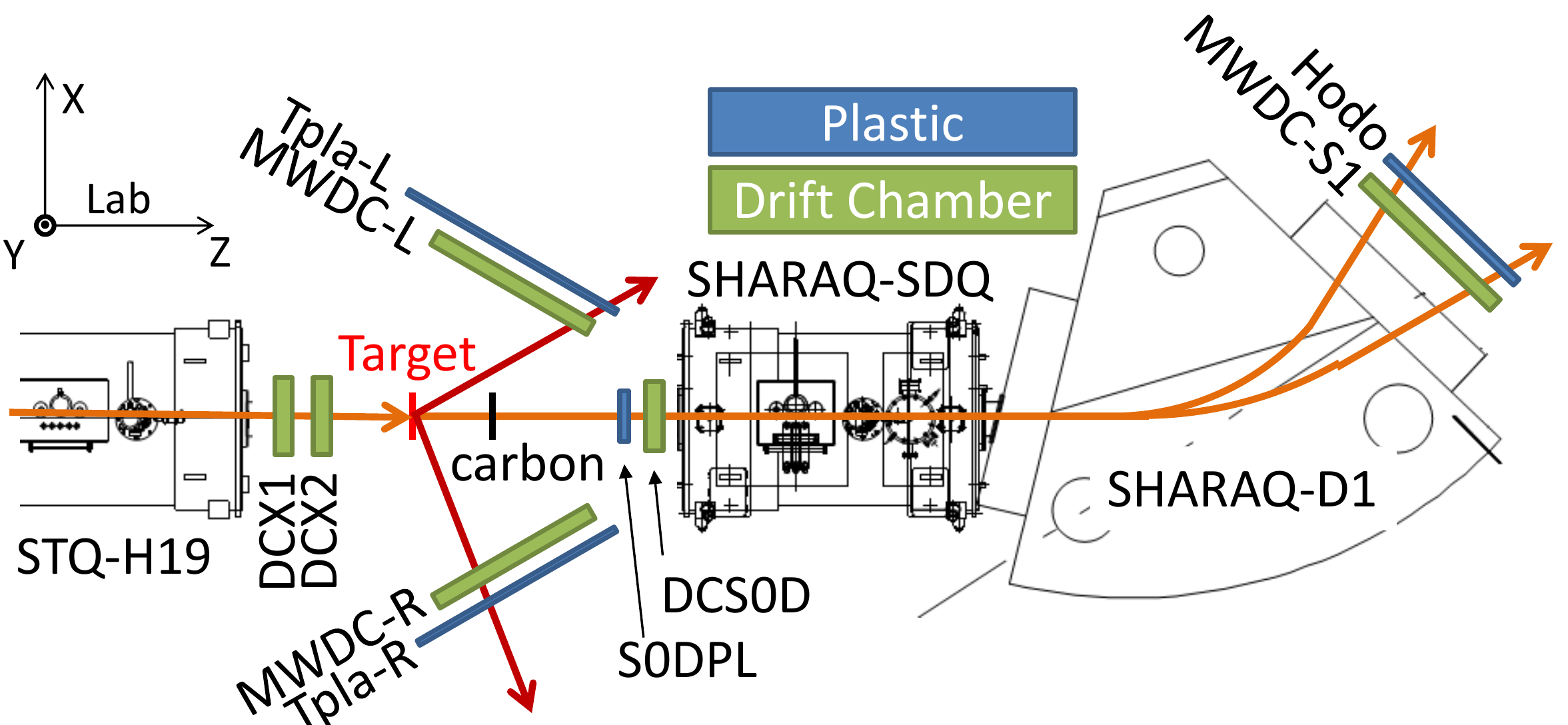}
\caption{Experimental set-up near the target. Beam (orange) comes from the left and is tracked. Scattered protons (red lines) are detected by plastic scintillators (Tpla-L/R) and tracked by multi-wire drift chambers (MWDC-L/R). Residual nuclei are analyzed by downstream detectors with SHARAQ-SDQ and -D1 magnets. See Ref.~\cite{Kawase18} for details.}
\label{fig1}
\end{figure}
%=======================================

The (\textit{p},2\textit{p}) quasi-free knockout experiments with exclusive measurements using \iso{25}{F} and \iso{23}{F} beams were performed at the Radioactive Isotopes Beam Factory operated by RIKEN Nishina Center and Center of Nuclear Study, The University of Tokyo. The primary beam of \iso{48}{Ca} at 345\textit{A} MeV with an intensity of \mbox{200 pnA} hit a 30-mm thick $^{9}$Be target, and then a secondary beam was produced and transported through the \mbox{BigRIPS}~\cite{Kubo03} to the SHARAQ spectrometer~\cite{Uesaka12}. The \mbox{BigRIPS} was tuned to provide a cocktail beam, which included \iso{25}{F}. The particle identification (PID) of the secondary beam was conducted using the $\Delta$E-TOF-B$\rho$ method. The intensity, energy, and purity of \iso{25}{F} were $1.69\times10^{4}$ pps, \mbox{277\textit{A} MeV}, and 42\%, respectively. A \iso{23}{F} cocktail beam was also produced and analyzed. The secondary target was a 1-mm thick C$_{10}$H$_{8}$ crystal~\cite{Tang12,Tang13}. FIG.~\ref{fig1} shows the detector setup around the target. Details can be found in Ref.~\cite{Kawase18}. 

The upstream PID, the proton-proton coincidence, the reaction vertex, and the residues PID were used to identify the reaction channel. The missing four-momentum of the residual \iso{24}{O} was reconstructed using 
\begin{equation}\label{eq:2}
P_O = P_F + P_T - P_1 - P_2,
\end{equation}
where $P$ are the four-momenta of the oxygen residues ($O$), the fluorine nucleus ($F$), the target proton ($T$), and the two scattered protons ($1$ and $2$). The excitation energy $E_x$ of the residual oxygen was then deduced by
\begin{equation}\label{eq:3}
E_x= m(P_O) - m_O,
\end{equation}
where $m(P_O)$ is the invariant mass of oxygen residues $P_O$, and $m_O$ is the mass of ground state oxygen. 

\begin{figure}[!hb]
\centering
\subfigure[]{
	\includegraphics[width=3.7cm,trim={14cm 2cm 1cm 5cm}, clip=true]{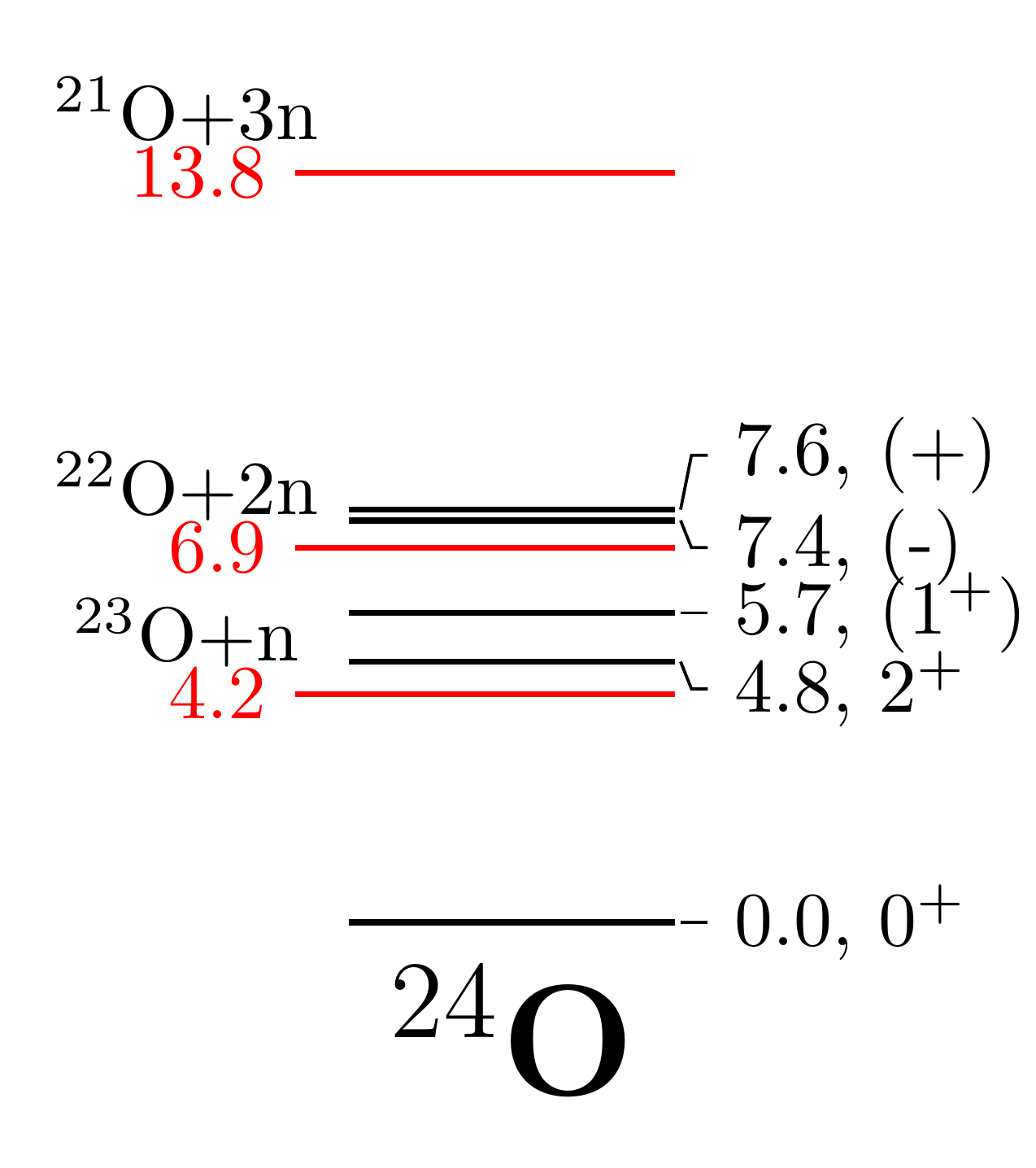}
	\label{fig2:A}
	}
\subfigure[]{
	\includegraphics[width=4.5cm,trim={0cm 0cm 0cm 0cm}, clip=true]{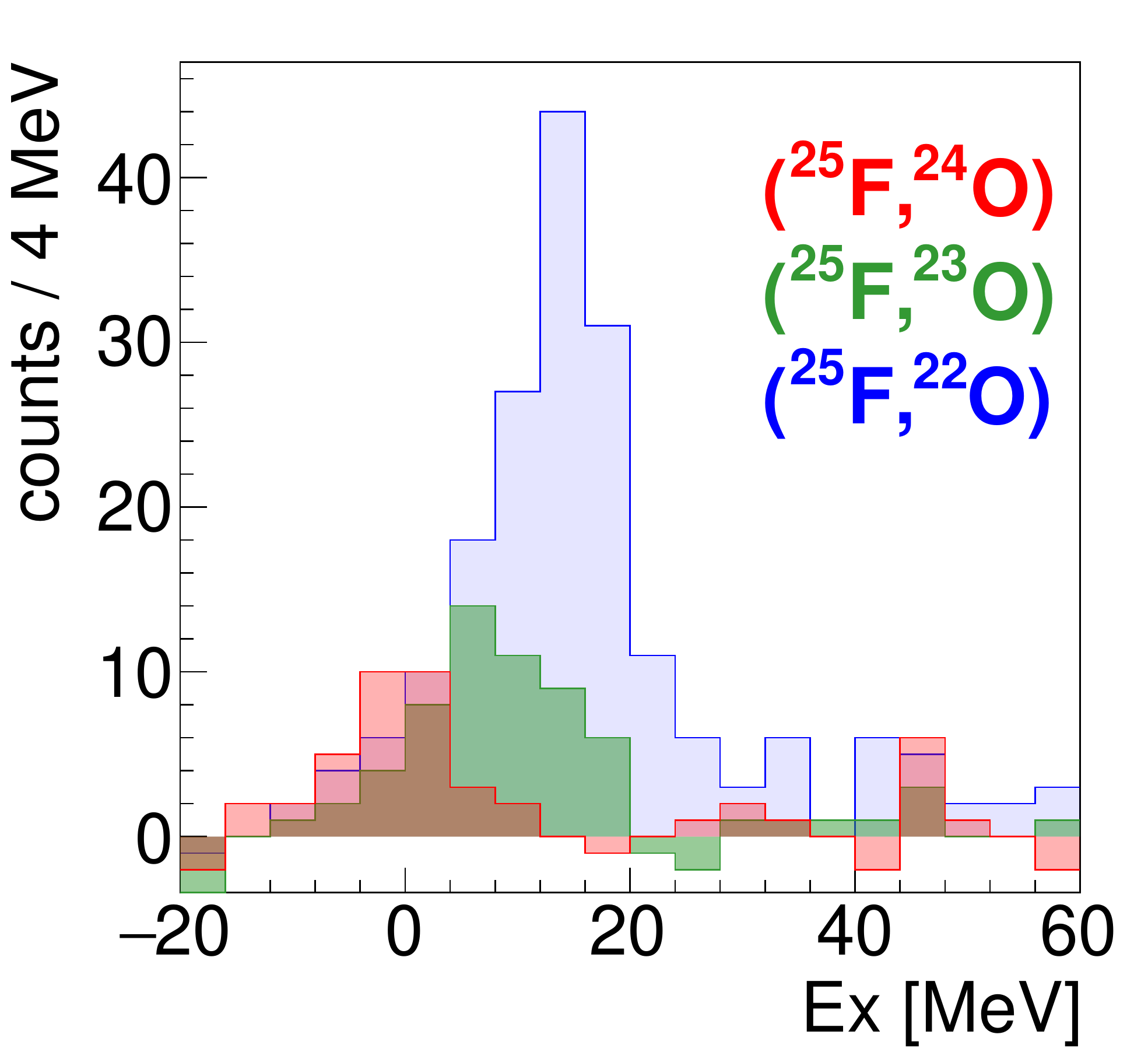}
	\label{fig2:B}
	}
\caption{
(a) Energy levels (black lines) and multiple-neutron threshold energies (red lines) of \iso{24}{O}~\cite{24O}. Unit of energy is MeV. (b) Experimental excitation-energy spectra of \iso{24}{O} from the \iso{25}{F}(\textit{p},2\textit{p}) reaction.}
\label{fig2}
\end{figure}

The residual oxygen nucleus can be excited above the neutron-emission thresholds and produce \iso{24-n}{O} isotopes. PID of the reaction fragments was used to select a group of excited states in \iso{24}{O} as shown in~[FIG.~\ref{fig2:A}]. In the following, the selection is notated as \ch{25}{24-n}. Data for \iso{23}{F} were analyzed using the same method. 

%===============  fig. 2  ==============
%======================================

%============ Table 1 ===================
\begin{table*}[ht]
	\caption{ Experimental results, integrated cross-sections, and spectroscopic factors. $S_{exp}$ extracted from experimental cross section and theoretical single-particle cross sections. $J^{\pi}_\textrm{th}$ is the spin-parity used for theoretical calculations.}
	\label{Table1}
	\centering
	\begin{tabular}{cccccccccccc}
\hline \hline
Channel & Mean [MeV] & Width [MeV]  & $\sigma_{\exp}$  [$\mu$b] & $\sigma_{\textrm{th}}$ [$\mu$b] & $J^{\pi}_\textrm{th}$ & $S_{\exp}$ &  $S_{\textrm{th}}$(USDB) &  $S_{\textrm{th}}$(SFO) & $S_{\textrm{th}}$(SPDF-MU)\\ 
\hline
\ch{25}{24} & -0.5(1.1) & 4.8(1.3)  & 53(18) & 149(24) & 5/2$^+$ & 0.36(13)& 1.01&0.90& 0.95 \\				
\ch{25}{23} &  6.5(1.4) & 6.3(9)    & 81(26)  &  125(26) & 5/2$^+$ & 0.65(25)& 0.01  &0.07 &  0.05   \\
\ch{25}{22} & 12.7(6)& 7.6(6)       & 274(71) & 80(24) & 1/2$^-$ & 3.1(1.3)&&2.19&\\
\hline
\ch{23}{22} & 1.0(8) & 6.0(6)      & 61(14) & 166(28) & 5/2$^+$ & 0.37(10)&1.08&0.92 & 1.00\\
\ch{23}{21} & 9.5(4) & 7.9(4)      & \multirow{2}{*}{456(67)} &  \multirow{2}{*}{93(25)}&  \multirow{2}{*}{1/2$^-$,3/2$^-$} &\multirow{2}{*}{4.9(1.5)}& & \multirow{2}{*}{5.21} &\\
\ch{23}{20} & 18.0(5)& 9.7(5)      &            &       & &&&\\
\hline \hline
	\end{tabular}
\end{table*}
%======================================

FIG.~\ref{fig2:B} shows the experimental excitation-energy spectrum of the \iso{25}{F}(\textit{p},2\textit{p}) reaction. Each \ch{25}{24-n} channel was fitted with a single Gaussian to obtain the mean excitation energy and the width.  The mean energy of the \ch{25}{24} channel is $-0.5\pm1.1$~MeV and is consistent with transition to the \iso{24}{O} ground state. The center of \ch{25}{23} channel is at \mbox{$6.5\pm1.4$~MeV}. The mean energy is located between the 1-neutron and the 2-neutron emission thresholds [FIG.~\ref{fig2:A}] as it should be. The mean energy of the \ch{25}{22} channel is $12.7\pm0.6$~MeV. The counts of \ch{25}{24}, \ch{25}{23}, and \ch{25}{22} channels were integrated from $-20$ to $20$ MeV, $-16$ to $26$ MeV, and $-8$ to $32$ MeV, respectively.~The cross section of each channel is obtained from dividing the integrated counts by the total luminosity, detector efficiency and acceptance.  \mbox{Table~\ref{Table1}} lists the experimental results along with that of \iso{23}{F}(\textit{p},2\textit{p}) reaction.

The orbital of the knocked-out proton for each channel can be identified by comparing the experimental proton momentum distribution to the DWIA calculation and the parity of the known states of the oxygen residues.~Additionally, the estimated shell-gap energy between $\pi$\orb{0}{d}{5} and $\pi$\orb{0}{p}{1} orbits could also support the orbital assignment. The momentum distribution of the knocked-out proton was reconstructed using the incident fluorine nucleus and the two scattered protons in the rest frame of the fluorine nucleus. 

To compare the experimental momentum distribution and extract the SF ($S_{\exp}$), DWIA calculations were conducted using the codes \textsc{PIKOE}~\cite{pik17} with a microscopic folding potential based on the Melbourne G-matrix interactions~\cite{Amo00} and nuclear densities calculated with the single particle potential by Bohr and Mottelson~\cite{BM69}. The calculations use the single proton wavefunction from the Woods-Saxon potential with a half-potential radius and a diffuseness parameter of $1.27A^{1/3}$ fm and \mbox{0.67 fm}, respectively~\cite{BM69}. These Wood-Saxon parameters is consistent with the optical potential~\cite{Kawase18}. The potential depth is calculated by matching the separation energy. A Coulomb radius of $1.25A^{1/3}$ fm is used. Another DWIA calculation is performed with code \textsc{THREEDEE}~\cite{Chant77} with the Dirac phenomenological potential with the parameters set EDAD2~\cite{Cooper93}. The results of both calculations agree within 5\%. 

The theoretical cross-section ($\sigma_{\textrm{th}}$) for a unit SF integrated over the detector acceptance is shown in \mbox{Table.~\ref{Table1}}, together with the experimentally integrated cross section ($\sigma_{\exp}$). The error of the DWIA-integrated cross-section was evaluated by taking the energy-dependence of the cross section and the uncertainty of the Wood-Saxon parameters into account. Finally, dividing the experimentally integrated cross-section by the theoretical integrated cross section, the SF ($S_{\exp}$) was obtained. 

%===============  fig. 3 ==============
% consider mono-color
\begin{figure}[!htb]
\centering
\subfigure[]{
	\includegraphics[width=4.1cm, trim={0.7cm 0cm 10.3cm 0cm}, clip=true]{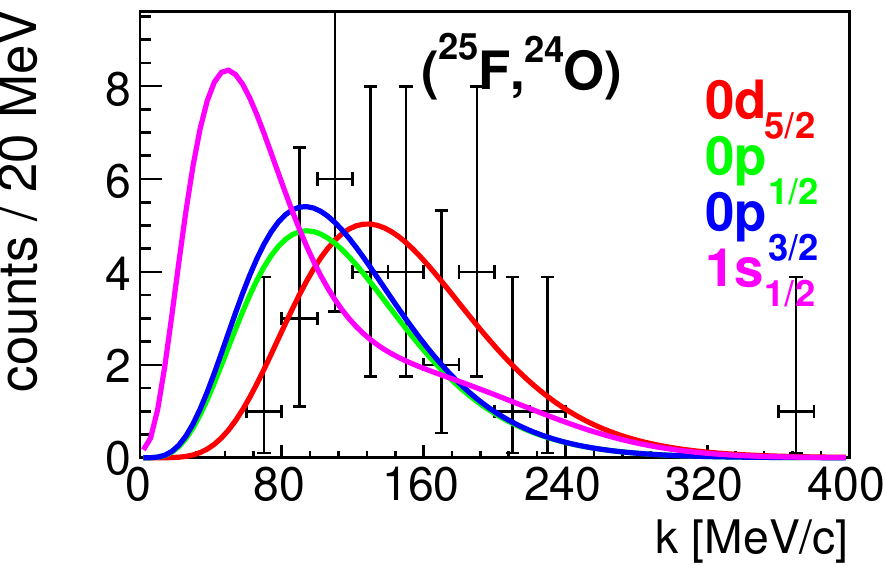}
	\label{fig3:A}
	}
\subfigure[]{
	\includegraphics[width=4.1cm, trim={10.7cm 0cm 0.3cm 0cm}, clip=true]{fig3.pdf}
	\label{fig3:B}
	}
\caption{Experimental proton momentum distribution of
different channels. Colored lines are the theoretical distributions. (a) \ch{25}{24} channel. (b) \ch{25}{23} channel.  }
\label{fig3}
\end{figure}
%======================================

\mbox{FIG.~\ref{fig3}} shows the experimental and theoretical proton momentum distributions of the \ch{25}{24} and \ch{25}{23} channels. Because the significant contribution of the s-orbital is not seen, one can consider that the protons are knocked out from the \orb{0}{d}{5} orbital in both channels. Since \iso{24}{O} does not have a bound excited state, this channel contains only the ground state. The parity of the \iso{24}{O} ground state is positive, and the momentum distribution [FIG.~\ref{fig3:A}] shows that the \ch{25}{24} channel is due to the knockout of \orb{0}{d}{5} proton. 

The momentum distribution for the \ch{25}{23} channel [FIG.~\ref{fig3:B}] is consistent with that for the knockout of \orb{0}{d}{5} orbital. Moreover, the parities of the known states are positive [FIG.~\ref{fig2:A}]. These support the idea that the channel is mainly due to knockout of \orb{0}{d}{5} proton. Considering the mean excitation energy for the \ch{25}{22} channel is \mbox{$12.7\pm0.6$ MeV}, which is comparable to the $\pi$\orb{0}{d}{5}--$\pi$\orb{0}{p}{1} shell gap of 12.7 MeV evaluated using proton separation energies of \iso{25}{F} and \iso{24}{O}, this channel is attributed to the \orb{0}{p}{1} proton knockout. 

%===============  fig. 4 ==============
\begin{figure}[!htb]
\centering
\includegraphics[width=8.5cm, trim={6.3cm 7.2cm 6.8cm 6cm}, clip=true]{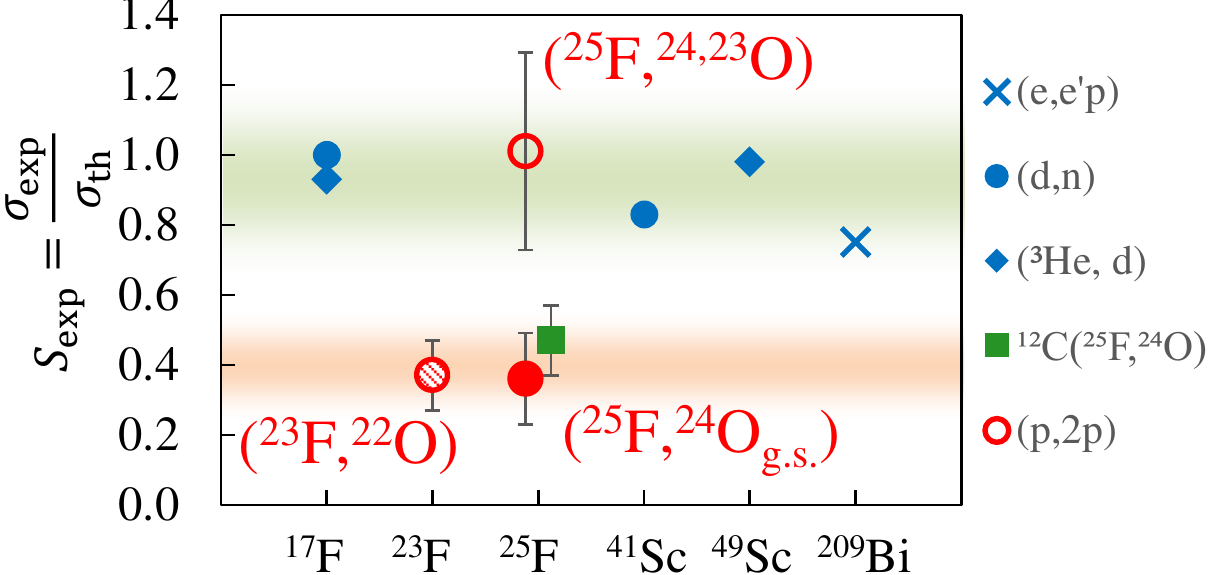}
\caption{Experimental SFs of various valence proton + doubly magic nuclei (\textcolor{lightblue}{$\times$}~\cite{Branford00}, \textcolor{lightblue}{$\bigcirc$}~\cite{Oliver69,Leighton68},  \textcolor{lightblue}{$\Diamond$}~\cite{Fortune75,Britton76}, \textcolor{green}{$\Box$}~\cite{Thoennessen03}, \textcolor{red}{$\bigcirc$} present work). Solid symbols represent the ground-state to ground-state channel. Open symbols represent the total strength of \orb{0}{d}{5} $\pi$-SPO. Shaded red circle is the sum of SFs of the \ch{23}{22} channel, which contains contributions from ground and excited states.}
\label{fig4}
\end{figure}
%\textcolor{lightblue}{$\triangle$}~\cite{Yasue92,Youngblood68,Gales84},
%======================================

The red circles in FIG.~\ref{fig4} show the SFs of the valence protons in \iso{23,25}{F}, where the solid circle denotes ground-state to ground-state transitions, the shaded circle is the sum of SFs from the observed d-orbital knockout channels, and the open circle is the estimated total strength from the d-orbital knockout. Blue symbol denotes the data of the ground-state to ground-state transition for other proton + doubly magic nuclei. It should be noted that the SFs are original values presented in the references and can be subject to change when modern shell model calculations and reaction analyses are applied. However, a small change of \mbox{$\sim0.2-0.3$} in the shown cases with large SFs, does not affect the following discussion. 

The SF of the \ch{25}{24} channel is found to be \mbox{$0.36\pm0.13$} which is significantly smaller than unity. The result is compatible with that of Ref.~\cite{Thoennessen03}(FIG.~\ref{fig4}).

The quenching of the \orb{0}{d}{5} proton is not solely responsible for the reduction of proton SF~\cite{Lapikas93}. The SF of $0.65\pm0.25$ for the \ch{25}{23} channel, which is due to the \orb{0}{d}{5} proton knockout, indicates that part of the proton \orb{0}{d}{5} strength is fragmented to the \iso{24}{O} excited states. The sum of the SFs of the \orb{0}{d}{5} proton from the ground state to the excited states is $1.0\pm0.3$. This value is close to that of the ground-state to ground-state transition for \iso{17}{F}, \iso{41}{Sc}, \iso{49}{Sc}, and \iso{209}{Bi}. Thus, it is reasonable that the \orb{0}{d}{5} proton is in a SPO, and the magicity at $Z=8$ is persistent in \iso{25}{F}.  

Considering the \orb{0}{d}{5} proton is in a SPO, the small overlap is due to the difference in the neutron shell structure~[Eq.~\eqref{eq:0}]. The small SF for the (\iso{25}{F}, \iso{24}{O}$_\textrm{g.s.}$) channel indicates the small overlap between the oxygen core of \iso{25}{F} and \iso{24}{O}$_\textrm{g.s.}$~[Eq.~\eqref{eq:0}]. In other words, the core of \iso{25}{F} would significantly differ from \iso{24}{O}$_\textrm{g.s.}$. The SFs show that only $\sim$35\% of the \iso{25}{F} core is \iso{24}{O}$_\textrm{g.s.}$ and the remaining $\sim$65\% is the \iso{24}{O} excited states. 

\begin{figure}
\centering
\subfigure[]{
	\includegraphics[width=7cm, trim={7cm 6.2cm 4.3cm 5cm}, clip=true]{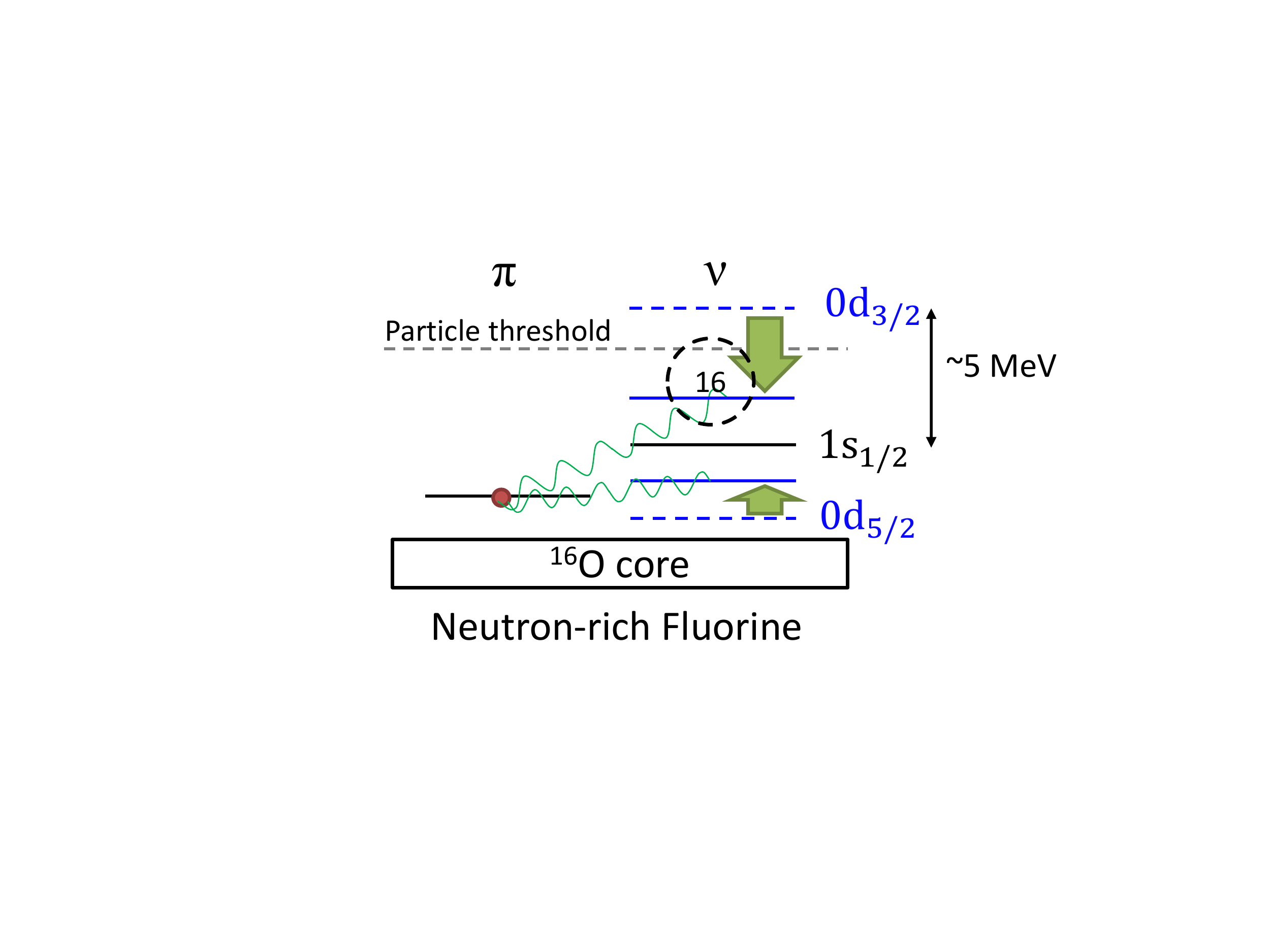}
	\label{fig5:A}
	}
\subfigure[]{
	\includegraphics[width=7cm,trim={7.2cm 6cm 6.6cm 5.5cm}, clip=true]{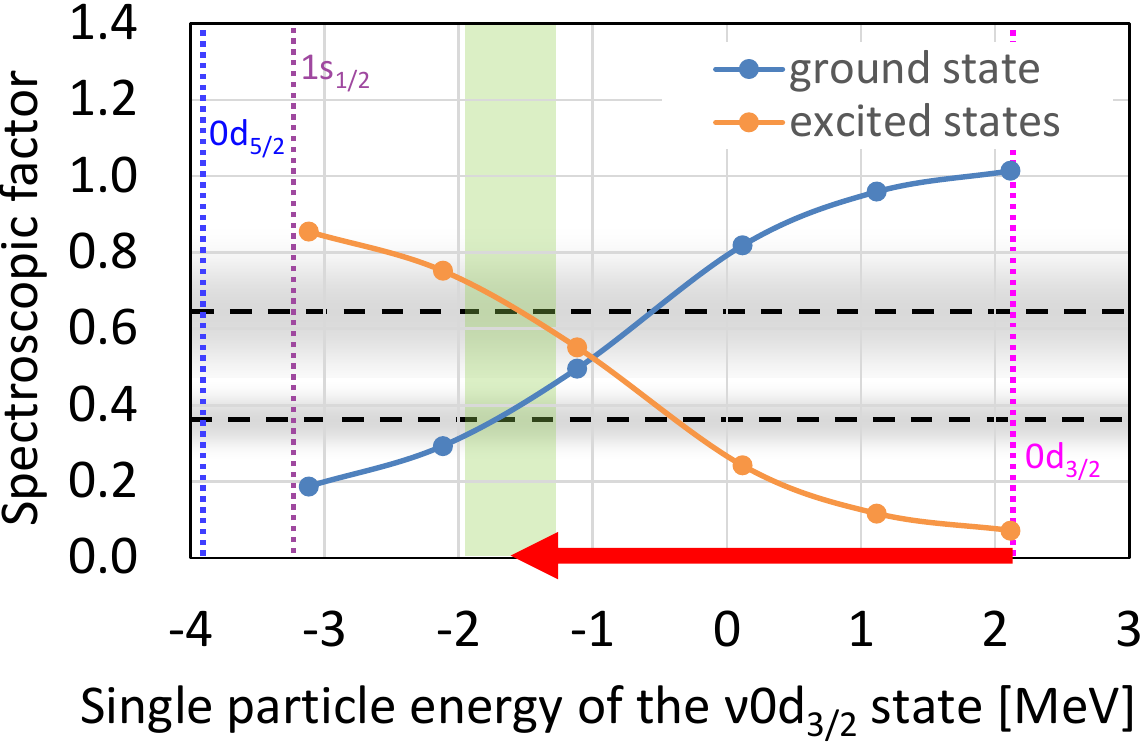}
	\label{fig5:B}
	}
\caption{(a) Mechanism of the Type-I shell evolution driven by the tensor force~\cite{Otsuka10, Otsuka05, Otsuka16}. Blue dashed lines are the energy levels of the d-orbits of oxygen. Blue solid lines are the energy levels of the d-orbits of fluorine. The wavy green lines represent the $pn$ tensor interaction. The \orb{1}{s}{1} orbital is assumed to be unaffected by the \orb{0}{d}{5} proton. The energy gap between \orb{0}{d}{3} and \orb{1}{s}{1} orbits in \iso{24}{O} is \mbox{$\sim$5 MeV}. (b) Modified shell model calculation based on USBD interaction. Vertical dotted lines are the original SPEs. Two black dashed lines and the gradients are SFs and their errors for \ch{25}{24} and \ch{25}{23} channels When the energy of \orb{0}{d}{3} SPO becomes lower, the ground state spectroscopic factor reduces. The red arrow indicates the reduction of the SPE of the \orb{0}{d}{5} orbital. See main text for detail. 
}
\label{fig5}
\end{figure}

A strong $pn$ tensor interaction induced by the \orb{0}{d}{5} proton in \iso{25}{F} [FIG.~\ref{fig5:A}] may be a plausible mechanisms for the change in the neutron-shell structure. A first-order effect of the tensor interaction by the \orb{0}{d}{5} proton [FIG.~\ref{fig5:A}]~\cite{Otsuka10, Otsuka05, Otsuka16} attracts the \orb{0}{d}{3} neutrons and repels the \orb{0}{d}{5} neutrons. A reduction in the $\nu$\orb{0}{d}{5}---$\nu$\orb{0}{d}{3} energy gap results in a larger configuration mixing among the neutron orbits, causing a change in the structure of the \iso{25}{F} core from \iso{24}{O}$_{\textrm{g.s.}}$. It should be noted that a recent $\gamma$-ray spectroscopy experiment revealed an additional 1.7~MeV $1/2^+$ level in \iso{25}{F}~\cite{Vajta14}. This may indicate the disappearance of $N=16$ magicity and be related to the configuration mixing.

%===============  fig. 5 ==============
%=======================================

%------------------------------------------------------------------
%\subsection{Comparison with shell model interactions}

Shell model calculations with $0,1,$ and $2~\hbar\omega$ excitation were carried out to clarify whether the SFs are explained by present nuclear theories. The calculations use the OXBASH~\cite{BrownOxbash} with the USDB (sd model space)~\cite{Brown06}, SFO (psd model space)~\cite{Suzuki03}, and SDPF-MU (sdpf model space)~\cite{Utsuno12} interactions. The results of the ground-state SF (Table~\ref{Table1}) are in the range of $0.9-1.0$, contradicting with the experimental results ($S_{exp}$). 

The contradiction could be attributed to nuclear correlations that are not fully taken into account in the shell model calculations employed above. One possibility is stronger tensor correlation in the region. Although the effective interactions including the tensor component are successfully reproduced in many experimental results, the lack of data in the very neutron-rich oxygen and fluorine region may limit the accuracy. Ref.~\cite{Vajta14} found that 60\% of \iso{25}{F} ground state is 1-particle state from N$^2$LO calculation using constrains from the latest measurement of the level scheme. This suggests the 3N force is important. 

As a sensitivity test for the possibility of a stronger $pn$ tensor interaction, shell model calculations based on USDB interaction with a modified $\nu$\orb{0}{d}{3} single-particle energy (SPE) have been explored. Since the $pn$ tensor interaction lowers the $\nu$\orb{0}{d}{3} SPE [FIG.~\ref{fig5:A}], the $\nu$\orb{0}{d}{3} SPE may be a suitable parameter to represent the tensor force strength. To reproduce the results of the present work, the $\nu$\orb{0}{d}{3} SPE must be lowered by \mbox{$3-4$~MeV} [FIG.~\ref{fig5:B}]. This suggests a stronger $pn$ tensor interaction in the structure of neutron-rich fluorine and oxygen isotopes. A more sophisticated self-consistent modification of all matrix elements would be needed to properly understand the effect of the tensor force, while out of the scope of the present experimental work.

A sudden change in the neutron-shell structure due to the \orb{0}{d}{5} proton may provide insight into the oxygen neutron dripline anomaly. The $pn$ tensor interaction lowers the energy of the $\nu$\orb{0}{d}{3} orbital. This may paves the path for a long fluorine neutron dripline and extend the dripline from $N=16$ for oxygen to $N=22$ for fluorine. The 3N force can also play a considerable role in describing the nuclear structures of fluorine isotopes~\cite{Vajta14,Soma14}. The present work suggests that the effects of the $pn$ tensor interaction in the widely used effective interactions may be too weak to reproduce the observed difference between the \iso{25}{F} core and \iso{24}{O}$_{\textrm{g.s.}}$. Together with the experiments discussed in Ref.~\cite{Brown01, Otsuka13, Otsuka10, Baumann12}, the mechanism on how the \orb{0}{d}{5} proton changes the neutron-shell structure may be revealed.

%*********************************************************************
%\section{Conclusions}
In conclusion, the neutron-shell structure of \iso{25}{F} is investigated using the (\textit{p},2\textit{p}) quasi-free knockout reaction at \mbox{270\textit{A} MeV} under inverse kinematics. The \orb{0}{d}{5} proton knockout from \iso{25}{F} populates the \iso{24}{O} ground state with a smaller probability than the \iso{24}{O} excited states. This result indicates that the oxygen core of \iso{25}{F} is considerably different from \iso{24}{O}$_\textrm{g.s.}$, and has a larger overlap with the excited states of \iso{24}{O}. The change in the neutron-shell structure due to the \orb{0}{d}{5} proton may be responsible for the small overlap between \iso{25}{F} and \iso{24}{O}$_\textrm{g.s}$. A comparison with the shell model calculations indicates that the USDB, SFO, and SFPD-MU interactions are insufficient to reproduce the present results. A stronger tensor force or other mechanism such as the 3N force effects, or both, might be need to explain the experimental results. More experimental and theoretical studies are necessary to clarify the mechanism for the change in the core of neutron-rich fluorine from ground-state of oxygen isotopes.

%*********************************************************************
%\section{Acknowledgments}
The authors thank the staffs at the RIKEN Nishina Center and Center for Nuclear Study, Tokyo University for operating the accelerators and ion source during the experiment. The first author (T.L.~Tang) is grateful for above institutes and all the support from Research Center of Nuclear Physics, Osaka University. This work was supported by the RIKEN Nishina Center (No. NP1006-SHARAQ04), Center for Nuclear Study, University of Tokyo, scholarship of Japanese Government (Monbukagakusho), and by Oak Ridge National Laboratory under Contract DE-AC05-00OR22725.\bigskip

%*********************************************************************


\begin{references}
\bibitem{Vijay97} V.~R.~Pandharipande \textit{et al.}, Rev. Mode. Phys. \textbf{69}, 981 (1997).
\bibitem{Caurier05} E.~Caurier \textit{et al.}, Rev. Mod. Phys. \textbf{77}, 427 (2005).
\bibitem{Glendenning04} N.~K.~Glendenning, \textit{Direct Nuclear Reaction} (World Scientific, 2004).
\bibitem{Kramer01} G.~J.~Kramer \textit{et al.}, Nucl. Phys. A \textbf{679}, 267 (2001).
\bibitem{Wakasa17} T.~Wakasa \textit{et al.}, Prog. Part. Nucl. Phys. \textbf{96}, 32 (2017).
\bibitem{Lapikas93} L.~Lapik$\acute{a}$s \textit{et al.}, Nucl. Phys. A \textbf{553}, 297 (1993).
\bibitem{Frauendorf14} S.~Frauendorf \textit{et al.}, Nucl. Phys. A \textbf{78}, 24 (2014).
\bibitem{Oliver69} C.~J.~Oliver \textit{et al.}, Nucl. Phys. A \textbf{127}, 567 (1969).
\bibitem{Fortune75} H.~T.~Fortune \textit{et al.}, Phys. Rev. C \textbf{12}, 1723 (1975).
\bibitem{Yasue92} M.~Yasue \textit{et al.}, Phys. Rev. C \textbf{46}, 1242 (1992).
\bibitem{Leighton68} H.~G.~Leighton \textit{et al.}, Nucl. Phys. A \textbf{109}, 218 (1968).
\bibitem{Youngblood68} D.~H.~Youngblood \textit{et al.}, Phys. Rev. C \textbf{2}, 477 (1970).
\bibitem{Britton76} R.~M.~Britton \textit{et al.}, Nucl. Phys. A \textbf{272}, 91 (1976).
\bibitem{Berthier86} B.~Berthier \textit{et al.}, Phys. Lett. B \textbf{182}, 15 (1986).
\bibitem{Gales84} S.~Gales \textit{et al.}, Phys. Rev. C \textbf{31}, 94 (1985).
\bibitem{Branford00} D.~Branford \textit{et al.}, Phys. Rev. C \textbf{63}, 014310 (2000).
\bibitem{Jacob66} G.~Jacob \textit{et al.}, Rev. Mod. Phys. \textbf{38}, 121 (1966).
\bibitem{McDonald86} W.~J.~McDonald \textit{et al.}, Nucl. Phys. A \textbf{456}, 577 (1986).
\bibitem{Yasuda10} Y.~Yasuda \textit{et al.}, Phys. Rev. C \textbf{81}, 044315 (2010).
\bibitem{Ozawa00} A.~Ozawa, Phys. Rev. Lett. \textbf{84}, 5493 (2000).
\bibitem{Dominguez11} B.~Fern\'{a}ndez-Dom\'{i}nguez, Phys. Rev. C \textbf{84}, 011301 (2011).
\bibitem{Nociforo09}C.~Nociforo \textit{et al.}, AIP Conf. Proc. {\bf1165}, 90 (2009).
\bibitem{Janssens09}R.~V.~F.~Janssens, Nature {\bf459}, 1069 (2009).
\bibitem{Kaneko04} K.~Kaneko \textit{et al.}, Phys. Rev. C \textbf{69}, 061302 (2004).
\bibitem{Thoennessen03} M.~Thoennessen \textit{et al.}, Phys. Rev. C \textbf{68}, 044318 (2003).
\bibitem{Kubo03}T.~Kubo \textit{et al.}, Nucl. Instrum. Meth. B {\bf204}, 97 (2003).
\bibitem{Uesaka12}T.~Uesaka \textit{et al.}, Prog. Theor. Exp. Phys. {\bf266}, 03C007 (2012).
\bibitem{Tang12}T.~L.~Tang \textit{et al.}, CNS annual report 2012, 59 (2013).
\bibitem{Tang13}T.~L.~Tang \textit{et al.}, CNS annual report 2013, 11 (2014).
\bibitem{Kawase18}S.~Kawase \textit{et al.}, Prog. Theor. Exp. Phys. 021D01 (2018).
\bibitem{Atar18}L.~Atar \textit{et al.}, Phys. Rev. Lett. \textbf{120} 052501 (2018).
\bibitem{24O} R.~B.~Firestone, Nuclear Data Sheets 108, 2319 (2007).
\bibitem{pik17} K.~Ogata, K.~Yoshida, and Y.~Chazono, computer code {\sc pikoe}
\bibitem{Amo00} K.~Amos \textit{et al.}, \textit{Advances in Nuclear Physics} (Plenum, New York, 2000) Vol. 25, p. 275.
\bibitem{BM69} A.~Bohr and B.~R.~Mottelson, \textit{ Nuclear Structure} (Benjamin, New York, 1969), Vol.~I.
\bibitem{Chant77} N.~S.~Chant \textit{et al.}, Phys. Rev. C {\bf15}, 57 (1977).
\bibitem{Cooper93} E.~D.~Cooper \textit{et al.}, Phys. Rev. C {\bf47}, 297 (1993).
\bibitem{Otsuka05} T.~Otsuka \textit{et al.}, Phys. Rev. Lett. \textbf{95}, 232502 (2005).
\bibitem{Otsuka10} T.~Otsuka \textit{et al.}, Phys. Rev. Lett. \textbf{105}, 032501 (2010).
\bibitem{Otsuka16} T.~Otsuka \textit{et al.}, J. Phys. G:Nucl. Part. Phys. \textbf{43}, 024009 (2016).
\bibitem{Vajta14} Zs.~Vajta \textit{et al.}, Phys. Rev. C \textbf{89}, 054323 (2014).
\bibitem{BrownOxbash} B~ A.~Brown \textit{et al.}, "MSU-NSCL report number 1289".
\bibitem{Brown06} B.~A.~Brown \textit{et al.}, Phys. Rev. C \textbf{74}, 034315 (2006).
\bibitem{Suzuki03} T.~Suzuki \textit{et al.}, Phys. Rev. C \textbf{67}, 044302 (2003). 
\bibitem{Utsuno12} Y.~A.~Utsuno \textit{et al.}, Phys. Rev. C \textbf{86}, 051301 (2012).
\bibitem{Soma14} V.~Som\`a \textit{et al.}, EPJ Web of Conferences \textbf{66}, 02005 (2014).
\bibitem{Otsuka13} T.~Otsuka, Phys. Scripta. \textbf{2013}, 014007 (2013).
\bibitem{Brown01} B.~A.~Brown, Prog. Part. Nucl. Phys.\textbf{47}, 517 (2001).
\bibitem{Baumann12} T.~Baumann \textit{et al.}, Rep. Prog. Phys. \textbf{75}, 036301 (2012).

\end{references}
\end{document}